\documentclass[11pt]{article}
\usepackage{graphicx}
\usepackage{pgfplots}

\usepackage[colorlinks=true,citecolor=blue]{hyperref}
\usepackage{natbib}
\usepackage{graphicx}
\usepackage{amsfonts}
\usepackage{amsmath}
\usepackage{amssymb}
\usepackage{url}
\usepackage{fancyhdr}
\usepackage{indentfirst}
\usepackage{enumerate}
\usepackage{titlesec}
\usepackage{amsthm}
\usepackage{multirow}
\usepackage{dsfont}
\usepackage[misc]{ifsym}
\usepackage{subfigure}
\usepackage{caption}

\theoremstyle{definition}

\newtheorem{example}{Example}

\theoremstyle{plain}
\newtheorem{theorem}{Theorem}
\newtheorem{lemma}{Lemma}
\newtheorem{proposition}{Proposition}
\newtheorem{corollary}{Corollary}
\theoremstyle{remark}
\newtheorem{remark}{Remark}

\usepackage{cases}

\theoremstyle{definition}

\def\P{\mathbb{P}}
\def\p{\mathbb{P}}
\def\E{\mathbb{E}}

\def\R{\mathbb{R}}

\def\X{\mathcal{X}}

\def\X{\mathcal{X}}

\def\d{\,\mathrm{d}}

\newcommand{\VaR}{\mathrm{VaR}}
\newcommand{\ES}{\mathrm{ES}}
\newcommand{\LES}{\mathrm{ES}^{-}}

\usepackage[onehalfspacing]{setspace}

\usepackage{bm}
\usepackage{tikz-qtree}
\usepackage{tikz}

\def\id{\mathds{1}}

\title{A reverse ES (CVaR) optimization formula}
\author{Yuanying Guan\thanks{Department of Mathematical Sciences and Department of Finance \& Real Estate, 
DePaul University, USA.
\Letter~{\url{yguan8@depaul.edu}}} \and Zhanyi Jiao\thanks{Department of Statistics and Actuarial Science, University of Waterloo,  Canada. \Letter~{\url{z27jiao@uwaterloo.ca}}} \and Ruodu Wang\thanks{Department of Statistics and Actuarial Science, University of Waterloo,  Canada. \Letter~{\url{wang@uwaterloo.ca}}}}
\date{\today}

\begin{document}
	\maketitle
	\begin{abstract}
 
The celebrated Expected Shortfall (ES, also known as TVaR and CVaR) optimization formula implies that ES at a fixed probability level is the minimum of a linear real function plus a scaled mean excess function. We establish a reverse ES optimization formula, which says that a mean excess function at any fixed threshold is the maximum of an ES curve minus a linear function. Despite being a simple result, this formula reveals elegant symmetries between the mean excess function and the ES curve, as well as their optimizers. The reverse ES optimization formula is closely related to  the Fenchel-Legendre transforms, and our formulas are generalized from ES to optimized certainty equivalents, a popular class of convex risk measures. We analyze worst-case values of the mean excess function under two popular settings of model uncertainty to illustrate the usefulness of the reverse ES optimization formula, and this is further demonstrated with an application using insurance datasets. 

	\end{abstract}


\section{Introduction}
The Value-at-Risk (VaR) and the Expected Shortfall (ES, also known as TVaR and
CVaR)   are the two most popular risk measures in banking and insurance, and they are widely employed in regulatory capital computation, decision making, performance analysis, and risk management.  In particular, ES is the standard risk measure in the current Basel Accords as well as the Swiss Solvency Test, and VaR is standard in the insurance regulatory framework of Solvency II.  
 For general treatments of VaR and ES in actuarial science and finance, see \cite{DDGK05}, \cite{KGDD08} and \cite{MFE15}. In this paper, we will stick to the term ``ES"  as used by \cite{BASEL19} for this risk measure, although the term ``CVaR" is   commonly found in  the optimization literature.

 An influential result on VaR and ES is an optimization formula obtained by \cite{RU00,RU02}, which is the main motivation for this paper. 
We first give the definition of VaR and ES. Let $\X$ be the set of integrable random variables in an atomless probability space $(\Omega,\mathcal F,\p)$.  
At a probability level $\alpha\in [0,1]$, VaR has two versions as the left- and right-quantiles.
For $X\in \X$,  define 
  \begin{equation}\label{eq:1} 
 \begin{array}{l}
  \VaR^{-}_\alpha(X) =  \inf \{t\in \R: \p(X\le t) \ge \alpha\}; \\
  \VaR^{+}_\alpha(X)  =  \inf \{t\in \R: \p(X\le t) >\alpha\}.
  \end{array} 
  \end{equation}
  By definition, $\VaR_0^{-}(X)=-\infty$ and $\VaR_1^+(X)=\infty$.   ES at   probability level $\alpha\in [0,1]$  is defined as
  $$\ES_\alpha(X)=\frac{1}{1-\alpha}\int_\alpha^1 \VaR^-_\beta(X)\d \beta,~~~X\in \X,~\alpha\in [0,1),$$
  and $\ES_1(X)=\VaR^-_1(X)$. 
 It is well known that an ES is a coherent risk measure (\cite{ADEH99}) and a convex risk measure (\cite{FS16}),
 and it admits an axiomatization based on portfolio diversification (\cite{WZ21}).
 Below, we first present the celebrated formula of \cite{RU02}.

\begin{theorem}[ES optimization formula]
\label{th:ru02} 
For  $X\in \X$ and $\alpha \in (0,1)$, it holds 
\begin{align}
\label{eq:old}   
  \ES_\alpha(X)  =    \min_{t\in  {\R}} \left\{t + \frac{1}{1-\alpha}\E[(X-t)_+] \right\},
  \end{align}
  and the set of minimizers for \eqref{eq:old} is $[\VaR^{-}_\alpha(X),\VaR_\alpha^+(X)] $.
  \end{theorem} 
   Theorem  \ref{th:ru02} has been a cornerstone of risk management since \cite{RU00,RU02} and \cite{P00}.  This result has been  tremendously useful in the optimization of ES (see \cite{RU13} for a review) and it has also been widely taught in actuarial science, see e.g., \citet[Section 2.4.3]{DDGK05} and \citet[Section 5.6]{KGDD08}.  
   Among other implications, this formula directly gives an elementary proof of subadditivity of ES; see \cite{EW15} for a comparison with six other proofs.

In this   paper, we establish a new optimization formula based on ES, which can be seen as a reverse formula to Theorem \ref{th:ru02}. 
This formula reveals nice symmetries between the ES curve and the mean excess   function, as we discuss in Section \ref{sec:discuss}. 
The mean excess loss, also known as the stop-loss premium, has a deep root in actuarial science (e.g., \cite{DG82}). 
In Section \ref{sec:app}, we apply  the new formula  to two popular settings of model uncertainty, one induced by   information of mean and a higher moment and the other induced by a Wasserstein ball. In both settings, the worst-case ES admits an explicit formula in the recent literature (\cite{PWW20, LMWW21}) whereas the worst-case mean excess   function does not.
Two insurance loss datasets are studied in Section \ref{sec:5} to illustrate the obtained results on the mean excess loss under model uncertainty induced by a Wasserstein ball. 
We present a few further technical results in Sections \ref{sec:OCE} and \ref{sec:FL}; more precisely, the reverse ES optimization formula are generalized to the class of optimized certainty equivalents introduced by \cite{BT07} (Section \ref{sec:OCE}),
and two related formulas are obtained via   Fenchel-Legendre transforms (Section \ref{sec:FL}). 
Section \ref{sec:conclusion} concludes the paper.
  
\section{A reverse ES optimization formula}

We start from the observation from Theorem \ref{th:ru02} that $\ES_\alpha(X)$  for a fixed $\alpha \in (0,1)$ can be obtained through taking the minimum of a function involving $\E[(X-t)_+]$ over $t \in \R$.
Having a mathematical symmetry in mind, a natural question is whether $\E[(X-t)_+]$ for a fixed $t\in \R$ can be obtained through taking the maximum of a function involving  $\ES_\alpha(X)$ over $\alpha \in [0,1]$. This leads to the reverse ES  optimization formula, the main result of this paper.
In what follows, we always use the convention $0\times x=0$ for $x\in [-\infty,\infty]$.

\begin{theorem}[Reverse ES optimization formula]
\label{th:new1}  For $X\in \X$ and $t\in \R$, it holds
\begin{align} \label{eq:new1} \E[(X-t)_+]  =  \max_{\alpha \in [0,1]}\left\{ (1-\alpha)\left( \ES_\alpha(X) -t \right) \right\},
 \end{align}
and the set of maximizers for \eqref{eq:new1} is $[\p(X<t), \p(X\le t)]$. 
 \end{theorem}
 
 To prove Theorem \ref{th:new1}, we first present a useful lemma 
 which collects some standard properties of quantiles, which are known to specialists on quantiles. We provide a self-contained short proof since we could not find this precise formulation in the literature.
 
\begin{lemma}
\label{le1}
For $\alpha \in [0,1]$ and any random variable $X $, the following statements hold:
\begin{itemize}
\item[(i)] $ \alpha > \P(X \leq t)$ if and only if $ \VaR_{\alpha}^{-}(X) > t$;
\item[(ii)] $ \alpha < \P(X < t)$ if and only if $ \VaR_{\alpha}^{+}(X) < t$.
\end{itemize}
\end{lemma}
\begin{remark}
  Lemma \ref{le1} can be equivalently stated as 
(i) $  \P(X \leq t)\ge \alpha $ if and only if $ \VaR_{\alpha}^{-}(X) \le  t$;
(ii) $   \P(X < t)\le \alpha $ if and only if $ \VaR_{\alpha}^{+}(X) \ge t$. 
\end{remark}
 \begin{proof}
To show (i), denote by
$ A_{\alpha} = \left\{ t\in \R: \P(X \leq t) \geq \alpha \right\}. $
Note that $A_\alpha$ is  closed   in $\R$ since $t\mapsto \p(X\le t)$ is upper semicontinuous. This gives $\VaR_{\alpha}^{-}(X) = \min A_{\alpha}.$
Hence,
$\alpha>\p(X\le t) \Longleftrightarrow t\not \in A_\alpha  \Longleftrightarrow  \VaR_{\alpha}^{-}(X) >t.$ 
To show (ii), denote by
$ B_{\alpha} = \left\{ t\in \R: \P(X < t) \leq \alpha \right\}$ which is  closed   in $\R$ since $t\mapsto \p(X< t)$ is lower semicontinuous. This gives $\VaR_{\alpha}^{+}(X) = \max B_{\alpha}.$
It follows that 
$\alpha<\p(X< t) \Longleftrightarrow t\not \in B_\alpha  \Longleftrightarrow  \VaR_{\alpha}^{+}(X) <t.$  
\end{proof}
\begin{proof}[Proof of Theorem \ref{th:new1}]
Let $g: [0,1] \to \mathbb{R}$, $ \alpha \mapsto (1-\alpha)(\ES_{\alpha}(X)-t) $.  
Note that for any $\alpha,\alpha'\in [0,1]$,
\begin{align}\label{eq:pf1}  
g(\alpha)-g(\alpha') &=  \int^{\alpha'}_{\alpha} \left(\VaR^{-}_{\beta}(X) -t\right)\d\beta 
\\&=    \int^{\alpha'}_{\alpha}\left(\VaR^{+}_{\beta}(X) -t\right) \d\beta . \label{eq:pf2} 
\end{align}
Let $[c,d]= [ \p(X < t), \p(X\le t)]$. 
For $\alpha\le  d$,    Lemma \ref{le1} (i) and \eqref{eq:pf1} imply 
$g(\alpha) \le  g(d)$.
For $\alpha < c$,   Lemma \ref{le1} (ii) and \eqref{eq:pf2} imply 
$g(\alpha) < g(c)$.
For $\alpha\ge c$,   Lemma \ref{le1} (ii) and \eqref{eq:pf2} imply  
$g(\alpha) \le g(c)$.
For $\alpha>d$,   Lemma \ref{le1} (i) and \eqref{eq:pf1} imply 
$g(\alpha) < g(d)$. 
Summarizing the above inequalities, we obtain
$$
g(\alpha_1 )<g(c) =g(\alpha_2) =g(d) > g(\alpha_3)\mbox{~~~~for any $\alpha_1<c<\alpha_2<d<\alpha_3$}.
$$
  Therefore, the set of maximizers for \eqref{eq:new1}
is  $[ c, d]$.  
By using Lemma \ref{le1} (i) again,
\begin{align*}
g(d) &= \int_{\p(X\le t)}^1\left(\VaR^{-}_{\beta}(X) -t\right) \d \beta 
= \int_{0}^1\left(\VaR^{-}_{\beta}(X) -t\right)_+ \d \beta 
=\E[(X-t)_+],
\end{align*}
thus showing \eqref{eq:new1}. 
\end{proof}


From the reverse ES optimization formula,  instead of directly calculating ${\E[(X-t)_{+}]}$ for fixed $t \in \R$, we can maximize a quantile-based function $\alpha \mapsto (1-\alpha)(\ES_{\alpha}(X)-t)$ over $\alpha \in [0,1]$. 
Some implications of this result are discussed in Section \ref{sec:discuss}.

The next corollary on a   formula for $ \E[ X\wedge x ]$ can be obtained from Theorem \ref{th:new1}.
To state this result, we define the left-ES for $\alpha \in [0,1]$ as
\begin{equation}\label{eq:2}
\LES_\alpha(X)=\frac{1}{\alpha}\int_0^\alpha \VaR^-_\beta(X)\d \beta,~~~X\in \X, ~\alpha\in (0,1],
\end{equation}
   and $\LES_0(X)=\VaR^+_0(X)$.

\begin{corollary}
\label{coro1}
For $t\in \R$ and $X\in \X$, it holds
\begin{align} \label{eq:new2} \E[ X\wedge t ]  = \min_{\alpha \in [0,1]}\left\{ \alpha \LES_\alpha(X)  +(1-\alpha)t   \right\},
\end{align}
and  the set of minimizers for \eqref{eq:new2} is $[\p(X<t), \p(X\le t)]$.
 \end{corollary}
 \begin{proof} 
The formula \eqref{eq:new1} directly leads to
\begin{align*}
\E [X \wedge t ]&= \E[X] - \max_{\alpha \in [0,1]} \left\{ (1-\alpha)(\ES_{\alpha}(X)-t)  \right\}\\
&=\E[X] + \min_{\alpha \in [0,1]} \left\{ (1-\alpha)(t - \ES_{\alpha}(X))  \right\}\\
&=\min_{\alpha \in [0,1]} \left\{ (1-\alpha)t +  \E[X]-(1-\alpha)\ES_{\alpha}(X)   \right\}\\ 
&=\min_{\alpha \in [0,1]} \left\{ (1-\alpha)t + \alpha\ES_{\alpha}^{-}(X)\right\}.
\end{align*} 
The corresponding statement on optimizers is the same as that in Theorem \ref{th:new1}.
\end{proof}
\begin{remark}
A similar relation to Corollary \ref{coro1} is found in \citet[p.15]{PR07} which is formulated for the   integrated distribution function and the integrated quantile function.
\end{remark}

\section{Symmetries between  Theorems \ref{th:ru02} and \ref{th:new1}}\label{sec:discuss}

The function $t\mapsto \E[(X-t)_+]$ is called the \emph{mean excess function} of $X$ according to \cite{MFE15}, 
and the function $\alpha\mapsto \ES_\alpha(X)$  
will be called the \emph{ES profile} of $X$ according to \cite{BMW22}.  
The ES profile also relates to the Lorenz curve (see e.g., \cite{G71}), which can be written as 
$ 
\alpha\mapsto  {\alpha\ES^-_\alpha(X)}/{\ES_0(X)}
$ 
for a non-negative random variable $X$ representing the wealth distribution of a population.
For clarity, we distinguish between the terms ``mean excess function" (as a function of $t$) and ``mean excess loss" (as a function  of $X$), and analogously between the terms ``ES profile" and ``ES".

 To appreciate Theorem \ref{th:new1} and contrast it with Theorem \ref{th:ru02}, we 
 need to understand the roles of the mean excess function and the ES profile. 
  The reason why Theorem \ref{th:new1} has not been explored in the literature is perhaps due to the perception that the ES profile is harder to obtain or to optimize than the mean excess function. 
  Based on this reasoning, it seems that using the mean excess function to compute ES is more natural than using the ES profile to compute the mean excess function.
 However,  in theory, there is no such asymmetry: For a given random variable $X$, its mean excess function and its ES profile have perfectly symmetric roles, as we discuss below.
 Indeed,  we shall see in Section \ref{sec:app} that in  relevant applications,  useful formulas for the mean excess function can be obtained  from the ES profile via Theorems \ref{th:new1}.

\begin{enumerate}
\item 
\textbf{Functional properties on $\X$.} Both the mean excess loss and ES have nice properties, symmetric to each other, as mappings  on $\X$.
 \begin{enumerate}[(a)]
\item For a  fixed $t\in \R$, the mapping $X\mapsto \E[(X-t)_+]$ is linear in the distribution of $X$ and convex in the quantile of $X$. Indeed, this mapping satisfies the independence axiom of \cite{vNM47}.
\item  For a  fixed $\alpha \in (0,1)$, the mapping $X\mapsto \ES_\alpha(X)$ is linear in the quantile of $X$ and concave in the distribution of $X$ (e.g., \cite{WWW20}). Indeed, this mapping satisfies the dual independence axiom of \cite{Y87}. 
\end{enumerate} 

\item \textbf{Optimization problems.} 
As for the optimization problems \eqref{eq:old} and \eqref{eq:new1}, we have the following symmetry. 
  \begin{enumerate}[(a)]
\item In the minimization \eqref{eq:old} over $t\in  {\R}$, the function $t\mapsto t + \frac{1}{1-\alpha}\E[(X-t)_+] $ is convex in $t$. 
  \item In the maximization \eqref{eq:new1} over $\alpha \in[0,1]$, the function $\alpha \mapsto 
  (1-\alpha)( \ES_\alpha(X) -t )$ is concave in $\alpha$. 
  \end{enumerate}
  
  \item \textbf{Solutions to the optimization problems.} 
  The optimizers to the optimization problems  \eqref{eq:old} and  \eqref{eq:new1}  also admit nice symmetry, as one is the quantile interval, and the other one is the probability interval.
  \item \textbf{Parametric forms.}  For commonly used distributions in risk management and actuarial science, if one of the mean excess loss and ES has an explicit formula, then so is the other one (e.g., Pareto distributions; see Example \ref{ex:pareto} below).
  Moreover, each of the two curves determines the whole distribution of the random variable.  
  \end{enumerate}
  
%
    
To summarize, writing one as a minimum or maximum of the other as in Theorems \ref{th:ru02} and \ref{th:new1} leads to the following implications for optimization:
\begin{enumerate}[(a)]
\item  Theorem \ref{th:ru02} allows one to transform the non-linear (in distribution)  objective $\ES_\alpha(X)$ as the minimum over $t$ of linear  (in distribution)   functions convex in $t$. 
\item  Theorem \ref{th:new1}  allows one to  transform the non-linear  (in quantile)  objective $ \E[(X-t)_+]$ as the maximum over $\alpha$ of linear (in quantile) functions concave in $\alpha$. 
\end{enumerate}

Due to the above  discussed  symmetries, Theorem \ref{th:new1} serves as a natural dual formula to Theorem \ref{th:ru02}.  
Indeed, Theorems \ref{th:ru02} and \ref{th:new1}  are closely related to Fenchel-Legendre transformations, which we will discuss in Section \ref{sec:FL}.

 \section{Worst-case risk under model uncertainty}
 \label{sec:app}
 
 As discussed in Section \ref{sec:discuss}, one of the greatest advantages of the ES optimization formula in Theorem \ref{th:ru02} is that it allows us to translate optimization problems of ES to those of the mean excess function.  More precisely, for a set of actions $A$ and  a loss function $f:A\times \R^d\to \R$,  Theorem \ref{th:ru02}  implies
 $$
 \min_{y\in A} \ES_\alpha(f(y,\mathbf X)) =  \min_{t\in  {\R}}\left\{t + \frac{1}{1-\alpha}  \min_{y\in A} \E[(f(y,\mathbf X)-t)_+] \right\},
 $$
 and thus, for the minimization of ES, it suffices to minimize the mean excess loss $\E[(f(y,\mathbf X)-t)_+]$ for each $t\in \R$, which is more convenient in many specific settings; see the review in \cite{RU13}. 
 
 In contrast, Theorem \ref{th:new1} has a maximum operator in its formula \eqref{eq:new1}, and it is useful in maximization problems. 
Moreover, even though risk often needs to be minimized,   a maximum naturally appears in the presence of model uncertainty, which is often addressed via a worst-case approach. The worst-case risk evaluation under uncertainty appears in, e.g., \cite{GS89} and \cite{MMR06} in the context of decision making,  \cite{GOO03} and \cite{ZF09}  in the context of   optimization, and \cite{EPR13}  in the context of   risk aggregation.  
More precisely, suppose that there is uncertainty about a random vector $\mathbf X$, assumed to be in a set $\mathcal U$, and    $f:\R^d\to \R$ is a loss function. 
 Theorem \ref{th:new1}  implies that the worst-case mean excess loss can be computed by (recall that the convention is $0\times \infty=0$), via exchanging the order of  two suprema, 
\begin{align}
 \sup_{\mathbf X\in \mathcal U} \E[(f(\mathbf X)-t)_+] = 
  \sup_{\alpha \in [0,1]}\left\{ (1-\alpha)\left(  \sup_{\mathbf X\in \mathcal U}\ES_\alpha(f(\mathbf X)) -t \right) \right\}, \label{eq:exchange-r0}
\end{align}
which allows us to use rich existing results on  worst-case ES.  
Moreover, the maximization over $\alpha \in [0,1]$ is attainable under a condition of uniform integrability, as we show below.

\begin{proposition}\label{prop:r1}
Let $\mathcal Y$ be a   set of random variables and $t\in \R$. If $\mathcal Y$ is uniformly integrable, then 
\begin{align}
 \sup_{Y\in \mathcal Y} \E[(Y-t)_+] 
=\max_{\alpha \in [0,1]}\left\{ (1-\alpha)\left(   \sup_{Y\in \mathcal Y} \ES_\alpha(Y ) -t \right) \right\}. \label{eq:exchange}
\end{align}
In particular, if there exists $p>1$ such that $\sup_{Y\in \mathcal Y}\E[|Y|^p]<\infty$, then $\mathcal Y$ is uniformly integrable and \eqref{eq:exchange} holds.
\end{proposition}

\begin{proof}
We first show that uniform integrability of $\mathcal Y$ implies, for any $\alpha \in [0,1]$,
\begin{align}\label{eq:r1-pf1}
\lim_{\alpha'\to \alpha}\sup_{Y\in \mathcal Y} \int_\alpha^{\alpha'} |\VaR_\beta(Y)| \d \beta \to 0,
\end{align}
where the limit is one-sided if $\alpha=0$ or $\alpha=1$.
Suppose that \eqref{eq:r1-pf1} does not hold, and without loss of generality we consider $\alpha'\downarrow \alpha$ (in this case, $\alpha<1$). It follows that there exists $m>0$ such that, for any $\epsilon   \in (0,1-\alpha)$, there exists $Y_\epsilon\in \mathcal Y$ satisfying $\int_\alpha^{\alpha+\epsilon} |\VaR_\beta(Y_\epsilon)| \d \beta >m$. 
 Since $\VaR_\alpha$ is monotone in $\alpha$, we have $$m<\int_\alpha^{\alpha+\epsilon} |\VaR_\beta(Y_\epsilon)| \d \beta \le \int_0^\epsilon |\VaR_\beta(Y_\epsilon)| \d \beta +\int_{1-\epsilon}^1 |\VaR_\beta(Y_\epsilon)| \d \beta. $$
 For any $K>0$, let $\epsilon>0$ be such that $4 K \epsilon<m$.
 It follows that 
 \begin{align*}
 \E\left [|Y_\epsilon|\id_{\{|Y_\epsilon|>K\}}\right] & \ge \E[(|Y_\epsilon|-K)_+]
 \\& =  \int_0^1 ( |\VaR_\beta(Y_\epsilon)| -K)_+\d \beta
 \\&
  \ge \int_0^\epsilon ( |\VaR_\beta(Y_\epsilon)| -K)_+\d \beta +\int_{1-\epsilon}^1 (|\VaR_\beta(Y_\epsilon)| -K)_+ \d \beta
  \\&  \ge \int_0^\epsilon |\VaR_\beta(Y_\epsilon)| \d \beta +\int_{1-\epsilon}^1 |\VaR_\beta(Y_\epsilon)| \d \beta -2\epsilon K  >m/2.
 \end{align*}
Hence, $\sup_{Y\in \mathcal Y} \E[|Y| \id_{\{|Y|>K\}}]>m/2$, contradicting uniform integrability.
Therefore, \eqref{eq:r1-pf1} holds.

Let $g: [0,1] \to \mathbb{R}$, $ \alpha \mapsto  \sup_{Y\in \mathcal Y} (1-\alpha)(\ES_{\alpha}(Y)-t) $.  
Note that for any $\alpha,\alpha'\in [0,1]$, using \eqref{eq:r1-pf1},
\begin{align*} 
|g(\alpha)-g(\alpha') |&=\left| \sup_{Y\in \mathcal Y}   \int_{0}^{\alpha'} \left(\VaR^{-}_{\beta}(Y) -t\right)\d\beta -  \sup_{Y\in \mathcal Y} \int_0^{\alpha} \left(\VaR^{-}_{\beta}(Y) -t\right)\d\beta \right|
\\&\le \left|  \sup_{Y\in \mathcal Y}     \int^{\alpha'}_{\alpha}\left(\VaR^{-}_{\beta}(Y) -t\right) \d\beta \right|
\le   \sup_{Y\in \mathcal Y}     \int^{\alpha'}_{\alpha}   |   \VaR^{-}_{\beta}(Y)     | \d\beta + |(\alpha'-\alpha)t|,
\end{align*}
which converges to $0$ as $\alpha'\to\alpha$. This shows that $g$ is continuous on $[0,1]$, and hence the maximum in \eqref{eq:exchange} is attained. 
The last statement that  boundedness of $\E[|Y|^p]$ implies uniformly integrability can be found in Exercise 5.5.1 of \cite{D10}.
\end{proof}

In two settings of uncertainty based on moment information and the Wasserstein metric which we study below, explicit formulas for the worst-case ES are available in \cite{PWW20} and \cite{LMWW21}, whereas the worst-case  mean excess loss does not have an explicit formula. In the popular case that  $f$ is a portfolio loss function (i.e., $f(\mathbf x)=\mathbf w^\top \mathbf x$ for some $\mathbf w\in \R^d$),  
the multi-dimensional uncertainty sets reduce to one-dimensional sets of the same type; see \citet[Section 6]{MWW22}. For this reason, we will focus on the one-dimensional uncertainty sets.

 \subsection{Uncertainty sets induced by  moment information}
We first study the uncertainty set induced by mean and a higher moment.  For $p>1$, $m\in \R$ and $v \ge 0$,
denote by $$\mathcal L^p(m,v)=\{X\in \X:\E[X]=m,~ \E[|X-m|^p]\le v^p\},$$
 that is, the set of all random variables with given mean $m$ and a $p$-th centralized absolute moment at most $v^p$. 
 We are interested in the worst-case value of a functional over $\mathcal L^p(m,v)$. The special case of this problem when $p=2$, i.e., the setting with mean and variance information, has been the most popular; see e.g., \cite{GOO03},    \cite{L18} and \cite{LCLW20} on various risk measures.
 
Let $\rho:\mathcal L^p\to \R$ be a mapping where $\mathcal L^p$ is the set of all random variables with finite $p$-th moment. 
Note that the problem of  $  \sup_{X\in \mathcal L^p(m,v)}
\rho(X)$ is better suited for $\rho=\ES_\alpha$ or some other risk measures than for the mean excess loss    $\rho:X\mapsto \E[(X-t)_+]$, because many risk measures, including $\VaR_\alpha$ and $\ES_\alpha$, satisfy  some simple properties which yield
 $$  \sup_{X\in \mathcal L^p(m,v)}
\rho(X) =  m+ v \sup_{X\in \mathcal L^p(0,1)}
\rho(X) .$$
Therefore, we can convert the original problem to an optimization over $ \mathcal L^p(0,1)$. Such a relationship does not hold for the mean excess loss    $\rho:X\mapsto \E[(X-t)_+]$. 

The problem of the worst-case  mean excess loss with moment conditions has a long history;  see e.g., \cite{DG82} in actuarial science and \cite{J77} in operations research. 
  \citet[Corollary 1]{PWW20}  obtained a closed-form formula  for the worst-case $\ES_\alpha$ over $\mathcal L^p(m,v)$, that is,
  \begin{equation}\label{eq:PWW20}
    \sup_{X\in \mathcal L^p(m,v)} \ES_\alpha(X) = m+ v \alpha (\alpha^p (1-\alpha)+ (1-\alpha)^p\alpha)^{-1/p}.
    \end{equation}
    In particular,  in case $p=2$,   it becomes
 $ 
    \sup_{X\in \mathcal L^2(m,v)} \ES_\alpha(X) = m+ v  { (\frac{\alpha}{1-\alpha} )^{1/2} }$. 
By exchanging the order of two suprema, the problem of  worst-case mean excess loss
 can be obtained by combining \eqref{eq:PWW20} and Theorem \ref{th:new1}. In what follows, we use the convention that $1/0=\infty$ and $1/\infty=0$.
  \begin{proposition}
  \label{prop1}
   For $p>1$, $m,t\in \R$ and $v \ge 0$, we have 
    \begin{align}
    \label{eq:MU1}
  \sup_{X\in \mathcal L^p(m,v)} \E[(X-t)_+]  
     = \max_{\alpha \in [0,1]}  \left\{(1-\alpha)(m-t)  +   v \left ( (1-\alpha)^{1-p}+ \alpha^{1-p}\right )^{-1/p} 
       \right\}.
           \end{align} 
  \end{proposition}
  \begin{proof} 
  The proposition follows directly from putting together 
  \eqref{eq:exchange} and \eqref{eq:PWW20}. The maximum is attainable because  the set  $\mathcal L^p(m,v)$ satisfies the condition in Proposition \ref{prop:r1}.
  \end{proof}
  
\begin{figure}[h]
\centering
\includegraphics[width=10cm,height=9cm]{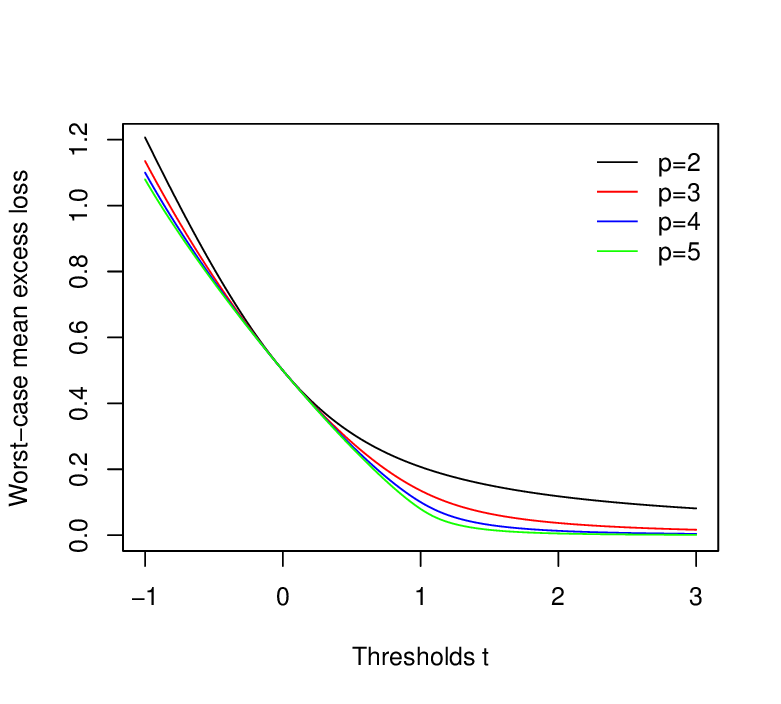}
\caption{Worst-case mean excess loss with moment conditions in $\mathcal L^p(0,1)$}\label{fig1}
\end{figure}

In the most popular case $p=2$, Proposition \ref{prop1} gives
    $$
  \sup_{X\in \mathcal L^2(m,v)} \E[(X-t)_+]  = \max_{\alpha \in [0,1]} \left\{ (1-\alpha)(m-t) + v\sqrt{\alpha(1-\alpha)} \right\}  = \frac{1}{2}\left( m-t + \sqrt{v^2+(m-t)^2} \right),
  $$
  which coincides with   \citet[Corollary 1.1]{J77}. The maximum value in \eqref{eq:MU1} for $p\ne 2$ can be   computed numerically.  
  We   provide a numerical example below by simply taking $m=0$ and $v=1$. Figure \ref{fig1} shows value of worst-case mean excess loss with respect to different thresholds $t$ under different moment conditions. We observe that a higher $p$  leads to a lower value of the worst-case mean excess loss at any threshold level, and this is because the constraint $\E[|X-m|^p]\le v^p$ is more stringent with larger $p$. 
   With Proposition \ref{prop1}, we can easily identify the worst-case values given a fixed threshold without knowing the exact distribution of loss. 

\subsection{Uncertainty sets induced by Wasserstein metrics}
Next, we consider uncertainty sets induced by Wasserstein metrics (this setting of uncertainty will be called the Wasserstein uncertainty).  Recall that the Wasserstein metric of order $p\ge 1$  between two distributions $F$ and $G$ on $\R$ is defined by 
$$ W_p(F,G) = \inf_{X\sim F, ~Y\sim G}\left(\E[|X-Y|^p]\right)^{1/p} =\left( \int^{1}_{0} \lvert F^{-1}(x) - G^{-1}(x) \rvert ^{p} \d x \right)^{1/p}, $$
where $X\sim F$ means that the distribution of $X$ is $F$.
For a benchmark loss $X\in \X=\mathcal L^1$ and an uncertainty level $\delta \ge 0$,   the \emph{Wasserstein ball} around $X$ is  
$ 
\left\{Y: W_p(F_X,F_Y) \leq \delta \right\},
$ where $F_X$ and $F_Y$ are the distributions of $X$ and $Y$, respectively. 
Note that $\delta=0$ corresponds to the case of no model uncertainty. 
The worst-case value of a risk measure $\rho:\X\to \R $ under the above uncertainty setting around $X$  is
$$   \sup \left\{ \rho(Y): W_p(F_X,F_Y) \leq \delta \right\}.$$ 
The  worst-case $\ES_{\alpha}$ under Wasserstein uncertainty is obtained by \citet[Proposition 4]{LMWW21} with the closed-form formula
\begin{equation}\label{eq:LMWW21}
 \sup \left\{ \ES_{\alpha}(Y): W_p(F_X,F_Y) \leq \delta \right\} = \ES_{\alpha}(X) + \frac{\delta}{(1-\alpha)^{1/p}}. 
    \end{equation} 
Based on \eqref{eq:LMWW21} and Theorem \ref{th:new1},  we can calculate the worst-case value of $\rho(X) = \E[(X-t)_{+}]$ for   $t\in \R$, similarly to Proposition \ref{prop1}. 

\begin{proposition}
  \label{prop2}
   For $t\in \R$, $p \geq 1$,  $\delta \ge 0$ and $X\in \X$, we have 
    \begin{equation}
   \label{eq:MU2}
  \sup \left\{ \E[(Y-t)_{+}]: W_p(F_X,F_Y) \leq \delta \right\} = \max_{\alpha \in [0,1]} \left\{ (1-\alpha) (\ES_{\alpha}(X)-t )+   {\delta(1-\alpha)^{1-1/p}}   \right\}.
  \end{equation}
  \end{proposition}
  \begin{proof}  The proposition follows directly from putting together 
  \eqref{eq:exchange-r0} and \eqref{eq:LMWW21}. The maximum is attainable because the function $\alpha\mapsto (1-\alpha) (\ES_{\alpha}(X)-t )+   {\delta(1-\alpha)^{1-1/p}}$ is continuous.
  \end{proof}
Comparing \eqref{eq:MU2} with  \eqref{eq:new1} in Theorem \ref{th:new1}, to compute the function $\E[(Y-t)_+]$ based on ES, there is an extra term of $\delta(1-  \alpha)^{1-1/p}$ in the maximization over $\alpha \in [0,1]$ to compensate for   model uncertainty.   As far as we are aware of, both formulas \eqref{eq:MU1} and \eqref{eq:MU2} in this section are  new. 
\begin{example}\label{ex:pareto}
Let the benchmark loss $X$ follow a Pareto distribution with tail parameter $\theta>1$, that is, $\p(X>x)= x^{- \theta}$ for $x \ge 1$.  For simplicity we take $\theta=2$ and   consider the Wasserstein metric $W_2$.
By straightforward calculation, $\ES_\alpha(X)= 2(1-\alpha)^{-1/2}$ for $\alpha \in [0,1)$. 
  Using \eqref{eq:MU2},    we get
  \begin{align} 
  \sup \left\{ \E[(Y-t)_{+}]: W_2(F_X,F_Y) \leq \delta \right\} \notag &= \max_{\alpha \in [0,1]} \left\{   \left(2+\delta\right) {  (1-\alpha)  ^{1/2}} - (1-\alpha) t   \right\} \\& =\frac{(1+\delta/2)^2}{t} \id_{\{t > 1+\delta/2\}} + (2+\delta -t)\id_{\{t  \le 1+\delta/2\}} .\label{eq:pareto}
  \end{align}
\end{example}
Example \ref{ex:pareto} also illustrates how the level of model uncertainty affects  the evaluation of the worst-case mean excess loss. Note that 
 for the benchmark loss $X$, 
  \begin{align}  \E[(X-t)_+] = \int_{t}^\infty \p(X>x) \d x = \int_{t\vee 1}^\infty x^{-2} \d x  +(1-t)_+ =  \frac 1t \id_{\{t > 1 \}} + (2 -t)\id_{\{t \le 1 \}}.  \label{eq:pareto2}
  \end{align} 
If $\delta=0$, then there is no model uncertainty, and  \eqref{eq:pareto} and \eqref{eq:pareto2} coincide. If $\delta>0$, then for $t>1 + \delta/2$,  
the worst-case value \eqref{eq:pareto} of the mean excess loss   increases from the benchmark value  \eqref{eq:pareto2}  by a factor of $(1+\delta/2)^2>1$;  for $t \le 1$, the  worst-case value \eqref{eq:pareto}  increases from the benchmark value  \eqref{eq:pareto2}  by a constant $\delta>0$.  We observe from \eqref{eq:LMWW21} that for $\ES_\alpha$ with a fixed $\alpha \in [0,1)$, the level of model uncertainty $\delta$ always affects the worst-case risk evaluation linearly; this also holds for any coherent distortion risk measures as shown by \cite{LMWW21}. 
In contrast, for the mean excess loss, the effect of $\delta$ is no longer linear in the interesting domain where $t$ is large.

\begin{example}\label{ex:normal}
Let the benchmark loss $X$ follow a normal distribution with mean $\mu$ and standard deviation $\sigma$. We can calculate $\ES_{\alpha}(X) = \mu + \sigma \frac{\phi(\Phi^{-1}(\alpha))}{1-\alpha}$ for $\alpha \in [0,1)$, where 
where $\phi$ and $\Phi^{-1}$ represent the density function and quantile function of the standard normal distribution, respectively. Using \eqref{eq:MU2}, we get 
\begin{equation*}
\sup \left\{ \E[(Y-t)_{+}]: W_p(F_X,F_Y) \leq \delta \right\} = \max_{\alpha \in [0,1]}\left\{ \sigma \phi(\Phi^{-1}(\alpha)) + (\mu-t)(1-\alpha) + \delta (1-\alpha)^{1-1/p} \right\}.
\end{equation*} 
Although the above expression is not explicit, it can be easily computed numerically. 
\end{example}

\begin{figure}[htp]
\centering
\subfigure[Worst-case value changes with $t$ (fixed $\delta = 0.1$)]{\includegraphics[width=8.2cm,height=7.9cm]{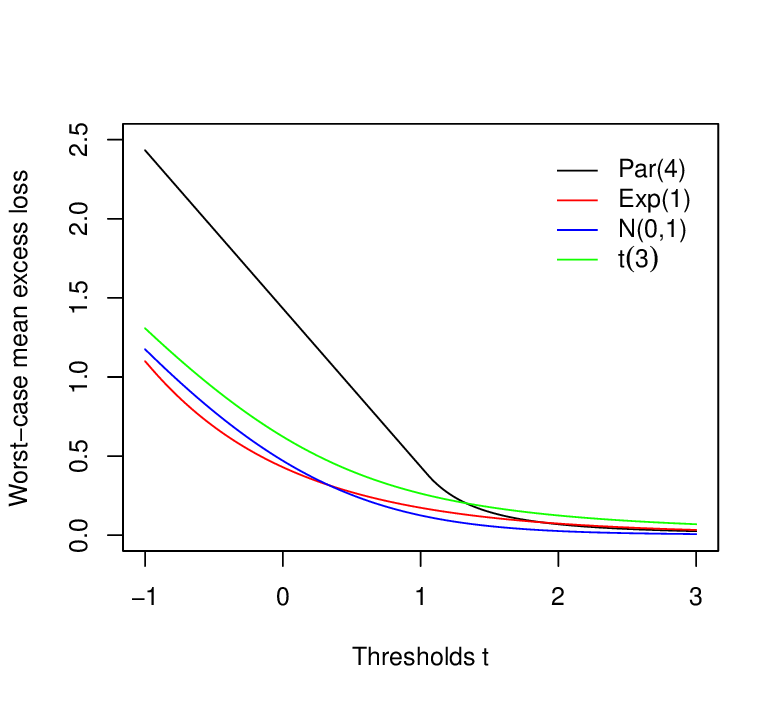}}
\subfigure[Worst-case value changes with $\delta$ (fixed $t = 2$) ]{\includegraphics[width=8.2cm,height=7.9cm]{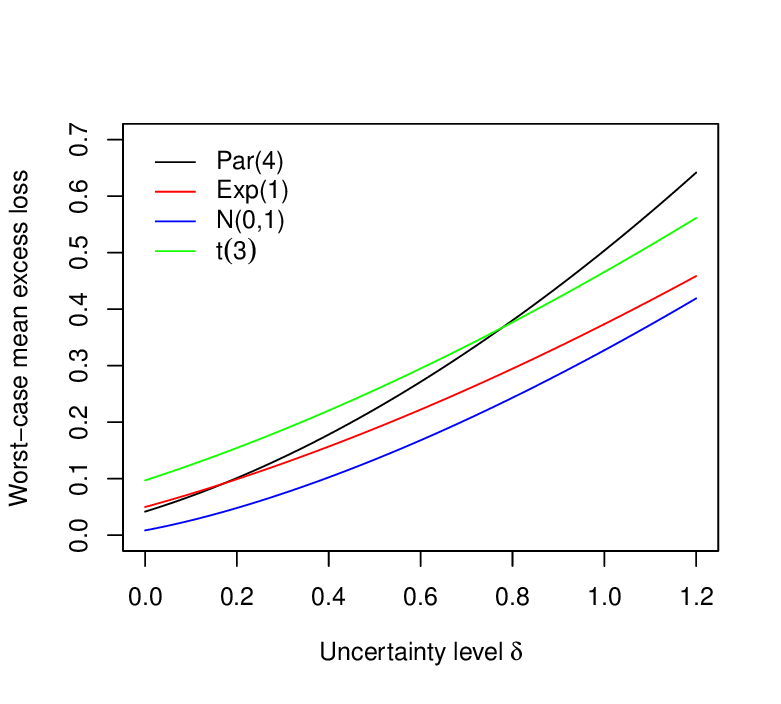}}\vspace{2.5ex}
\caption{Worst-case mean excess loss with Wasserstein uncertainty} \label{fig2}
\end{figure}

For a better understanding of Proposition \ref{prop2}, we provide a numerical example. In Figure \ref{fig2} (a), by choosing the parameter $p=2$ and the uncertainty level $\delta = 0.1$, we show how the worst-case values of the mean excess loss vary with the threshold $t$ under different distributions, including  Pareto, exponential, normal  and   t distributions. The obtained curves are similar   to  those in Figure \ref{fig1}. In
Figure \ref{fig2} (b), by taking $p=2$ and $t = 2$, we report the worst-case values of the mean excess loss increases with the uncertainty level $\delta$. As we can see,  the effect of $\delta$ on the worst-case value of mean excess loss is non-linear,  as we discussed in Example \ref{ex:pareto} for a Pareto distribution.


\section{Empirical analysis for insurance data}\label{sec:5}
In this section, we use  insurance data to calculate the worst-case mean excess loss under uncertainty   governed by the Wasserstein metric with $p=2$. In addition, we check how  the uncertainty level $\delta$ and the threshold level $t$   may influence the value of worst-case mean excess loss compared to the one without uncertainty, and see their different performances in different datasets. 

We choose two univariate datasets from  the  \texttt{R} package \texttt{CASdatasets}: normalized hurricane damages (\texttt{ushurricane}, 1900-2005) and normalized French commercial fire losses (\texttt{frecomfire}, 1982-1996) pooled by each month. Both  datasets have around 180 observations and the distributions are highly right-skewed. We shall use \verb!R! to fit the data with lognormal, Gamma and Weibull distributions as our benchmark distributions. 


In the first part of the empirical analysis, we fix the threshold level $t$ and let the uncertainty level $\delta$ vary to visualize how the worst-case mean excess loss varies. Since the two datasets are quite different, it is important to calibrate $t$ and $\delta$ to make two datasets to be comparable. In particular,  $\delta$ should be chosen in a statistically relevant range; see e.g.,  \cite{BCZ21} for a discussion  on this point. 
Generally, if the uncertainty level $\delta$ is too large, then the data become less relevant; if $\delta$ is too small, we are not protected against model uncertainty, thus losing the desired robustness. For a meaningful comparison, we make the following heuristic choices. For each benchmark distribution, we let $\delta$ vary in $ [\delta_0,2\delta_0]$, where $\delta_0$ is the Wasserstein distance (metric) between the fitted distribution and the empirical distribution.
This choice ensures that the empirical distribution is inside the Wasserstein ball around the fitted distribution; intuitively, a poorly fitted distribution is associated with a larger $\delta_0$, thus requiring a higher uncertainty level to be considered as robust.  Moreover, $\delta_0$ is of the order $n^{-1/2}$ if the estimation is $n^{-1/2}$-efficient, where $n$ is the sample size. 
Table \ref{tab:delta0} shows the values of $\delta_0$.

\begin{table}[t]
\begin{center}
\begin{tabular}{c|ccc}
\hline
\hline
& Lognormal & Weibull & Gamma \\
\hline
Hurricane loss & 43.83 & 37.99 & 47.45 \\
Fire loss & 186.69 & 248.99 & 244.24 \\
\hline
\hline
\end{tabular}
\end{center}
\caption{Values of $\delta_0$ for the lognormal, Weibull and Gamma distributions  and for the hurricane loss and fire loss datasets. The level $\delta_0$ is the Wasserstein metric with $p=2$ between the empirical and the fitted distributions}\label{tab:delta0}
\end{table}

\begin{figure}[h]
\centering
\subfigure[Empirical and fitted CDF]{\includegraphics[width=5.3cm,height=5cm]{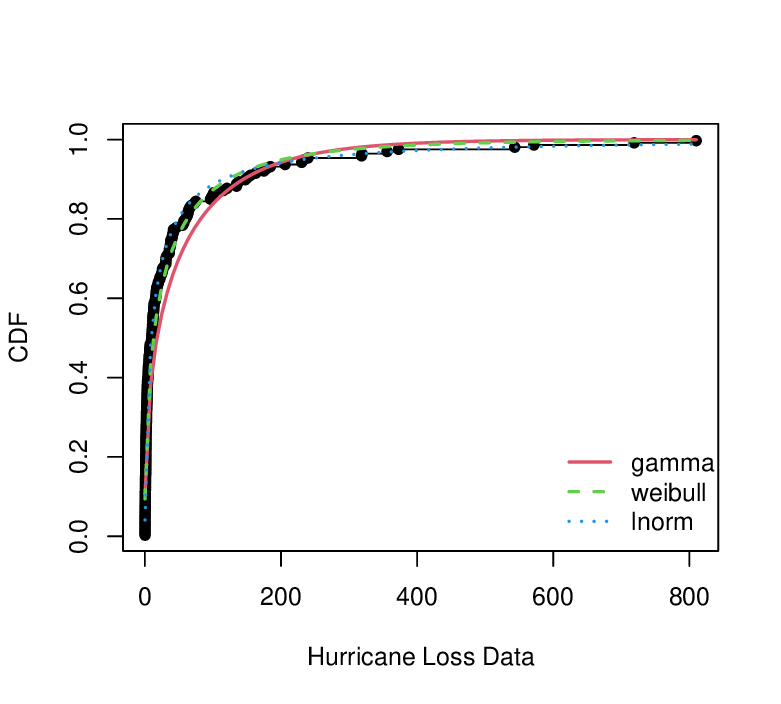}}
\subfigure[Q-Q plot]{\includegraphics[width=5.3cm,height=5cm]{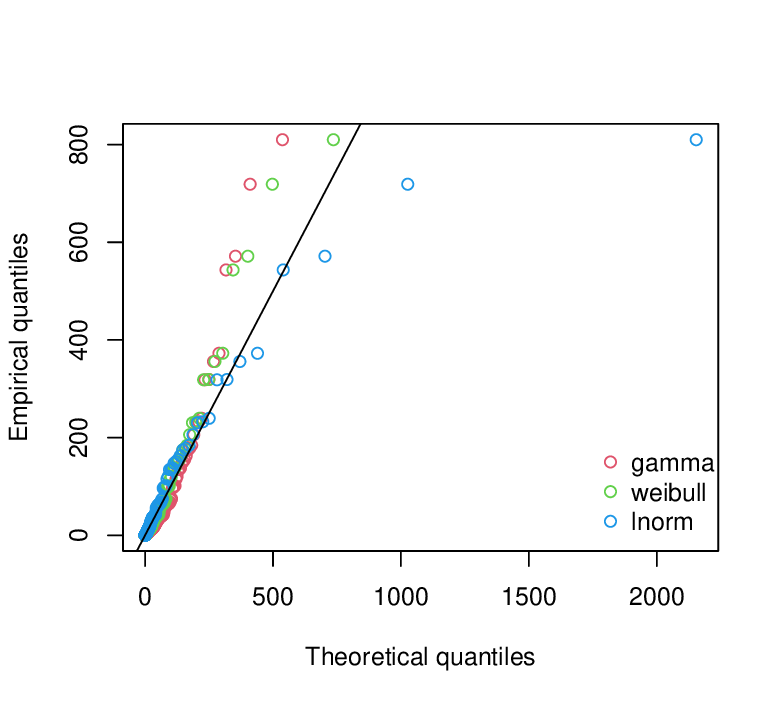}}
\subfigure[The curve of $r(\delta,t_0)$]{\includegraphics[width=5.3cm,height=5cm]{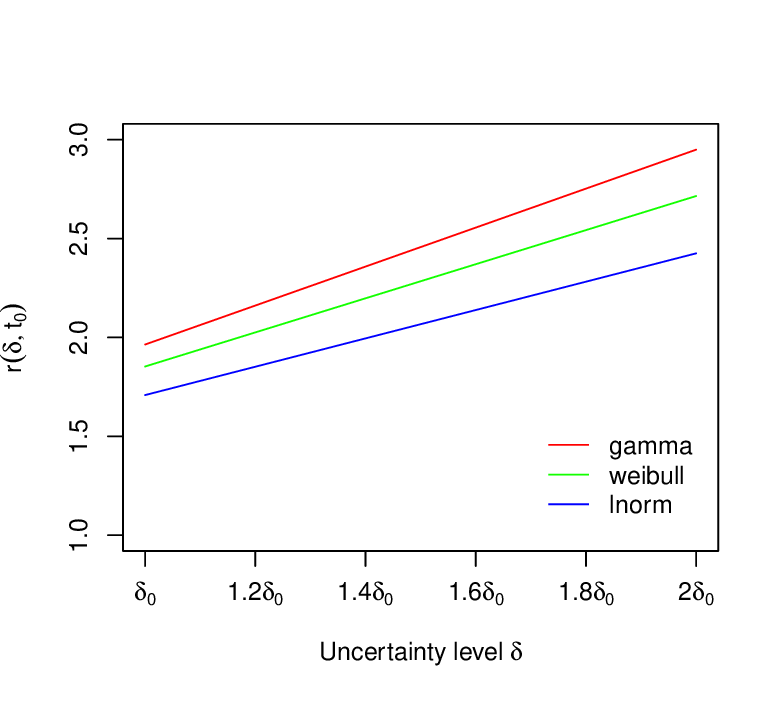}}
\vspace{2.5ex}
\caption{Empirical results on the hurricane loss data} \label{fig3}
\end{figure}

\begin{figure}[h]
\centering
\subfigure[Empirical and fitted CDF]{\includegraphics[width=5.3cm,height=5cm]{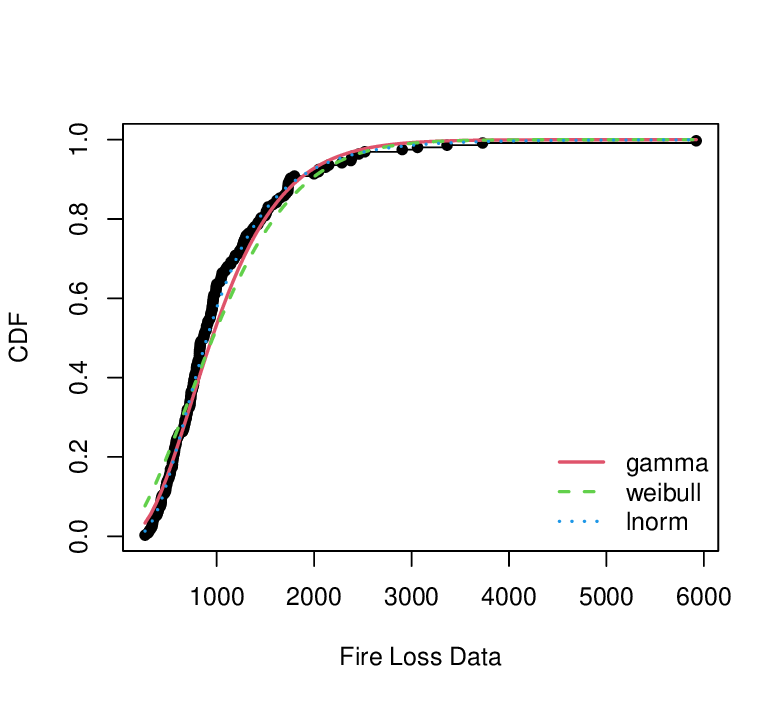}}
\subfigure[Q-Q plot]{\includegraphics[width=5.3cm,height=5cm]{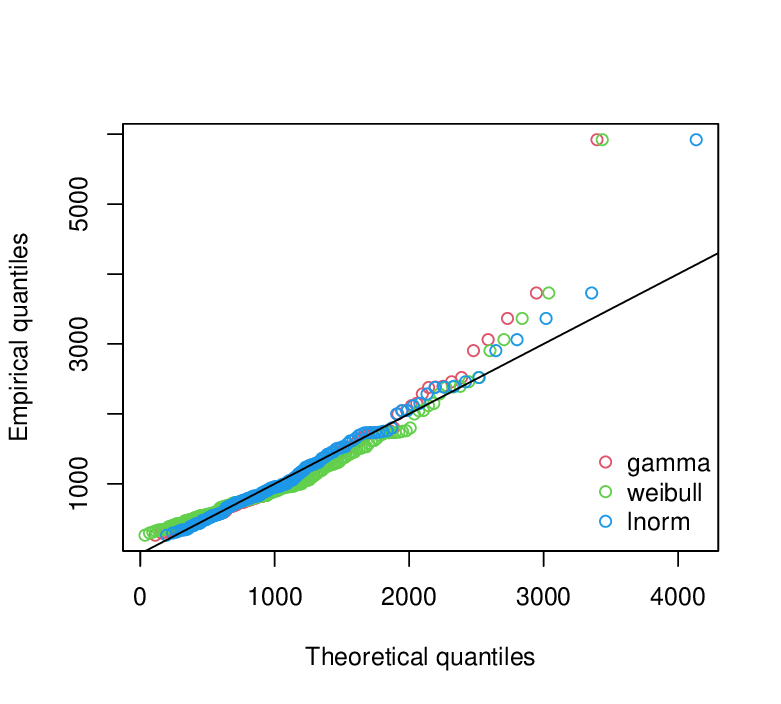}}
\subfigure[The curve of $r(\delta,t_0)$]{\includegraphics[width=5.3cm,height=5cm]{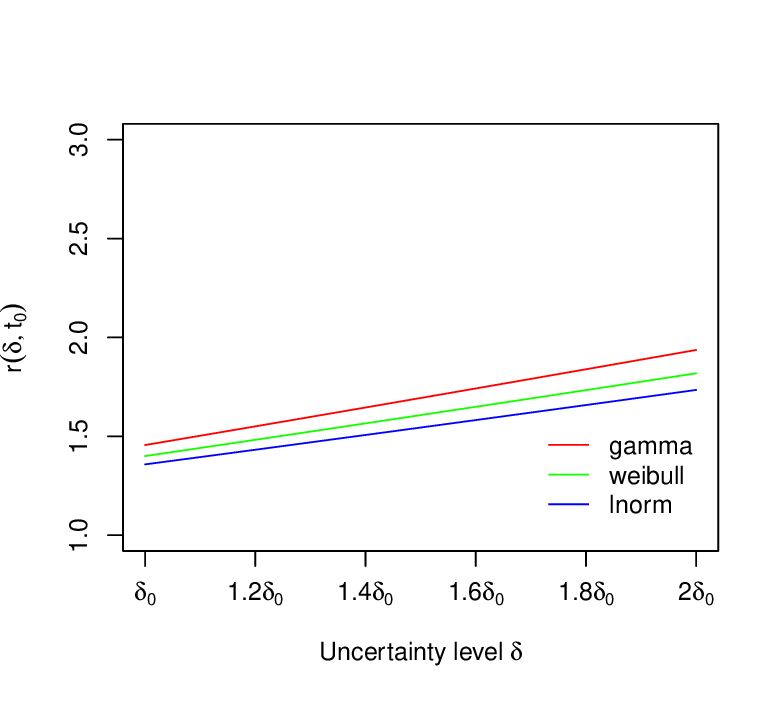}}
\vspace{2.5ex}
\caption{Empirical results on the  fire loss data} \label{fig4}
\end{figure}

\begin{figure}[ht!]
\centering
\subfigure[Fitted to a lognormal distribution]{\includegraphics[width=5.3cm,height=5.3cm]{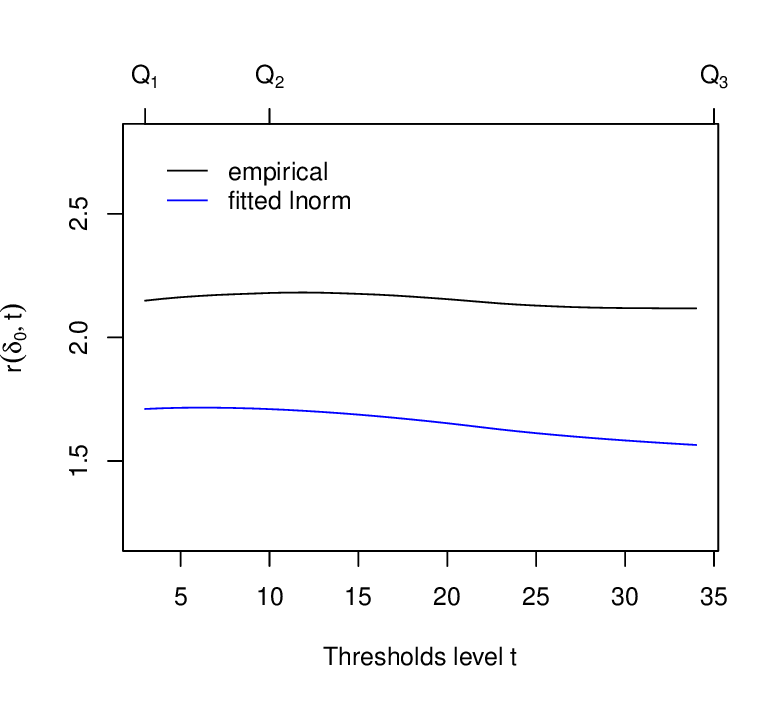}}
\subfigure[Fitted to a Weibull distribution]{\includegraphics[width=5.3cm,height=5.3cm]{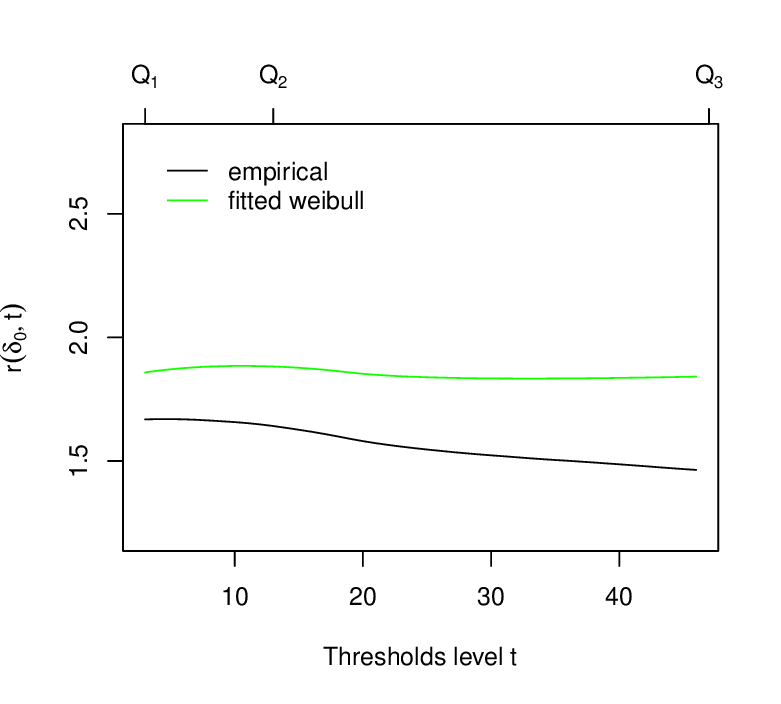}}
\subfigure[Fitted  to a Gamma distribution]{\includegraphics[width=5.3cm,height=5.3cm]{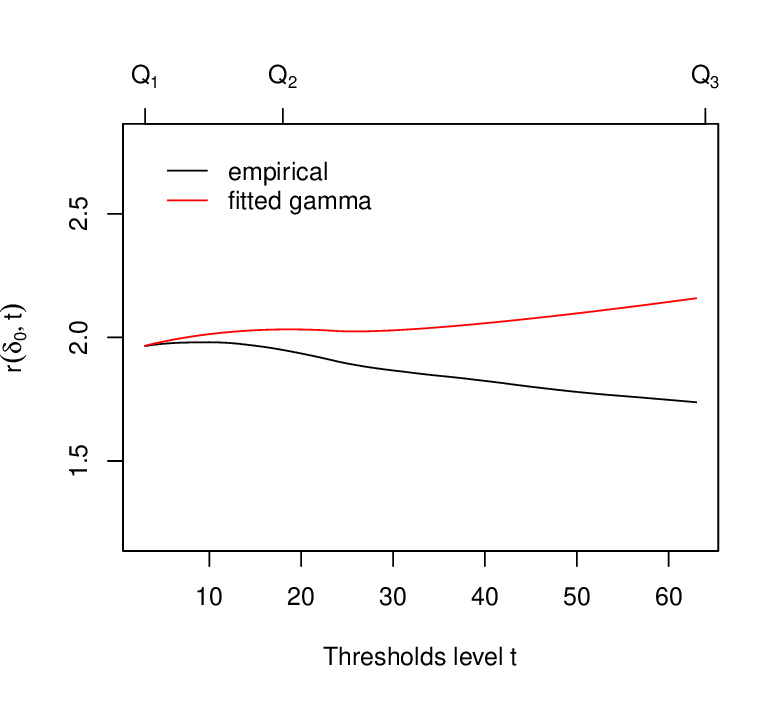}}
\vspace{2.5ex}
\caption{Values of the ratios  $r(\delta_0,t)$ and $\hat r(\delta_0,t)$ for the hurricane loss data, where $Q_1$, $Q_2$ and $Q_3$ represent the $1$st, $2$nd and $3$rd quartiles of the data} \label{fig5}
\end{figure}

\begin{figure}[ht!]
\centering
\subfigure[Fitted to a lognormal distribution]{\includegraphics[width=5.3cm,height=5.3cm]{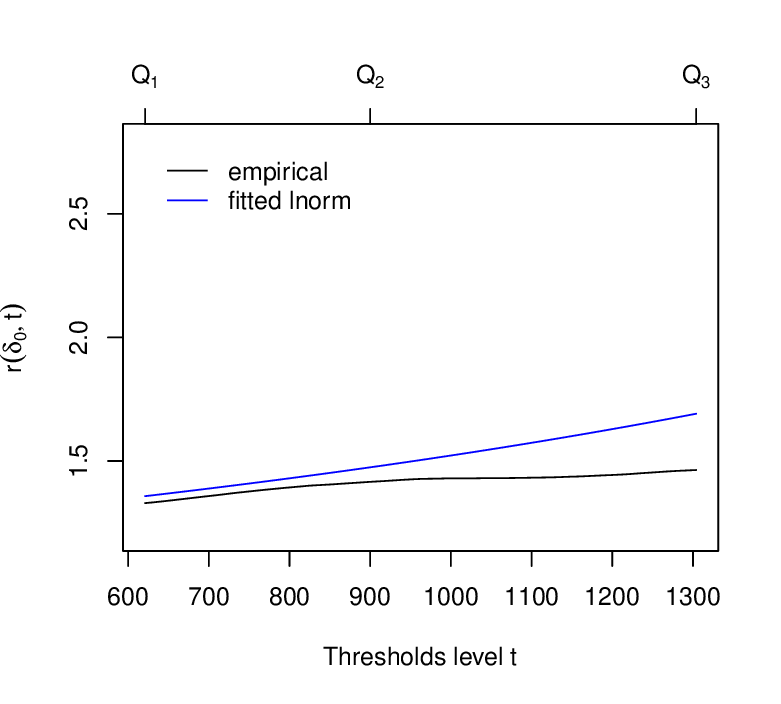}}
\subfigure[Fitted to a Weibull distribution]{\includegraphics[width=5.3cm,height=5.3cm]{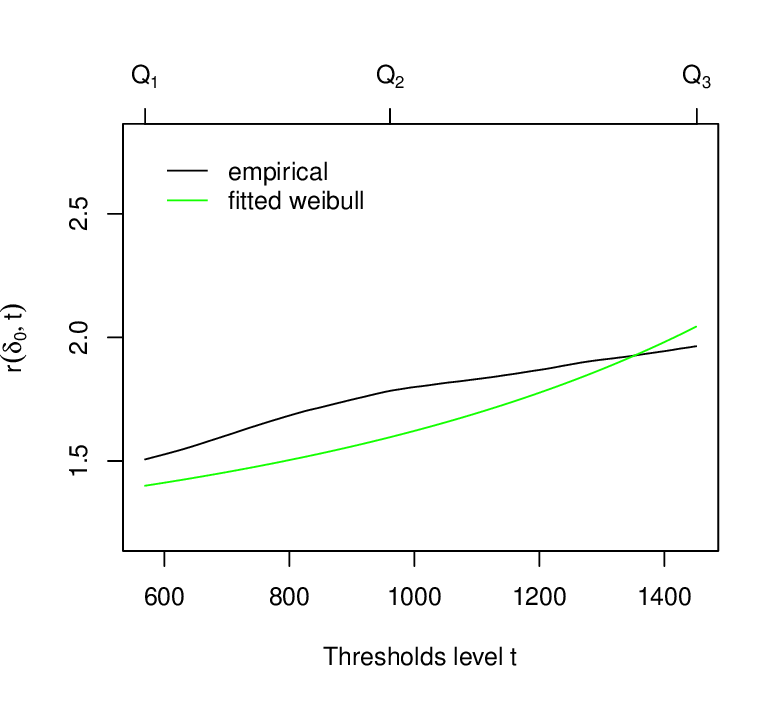}}
\subfigure[Fitted to a Gamma distribution]{\includegraphics[width=5.3cm,height=5.3cm]{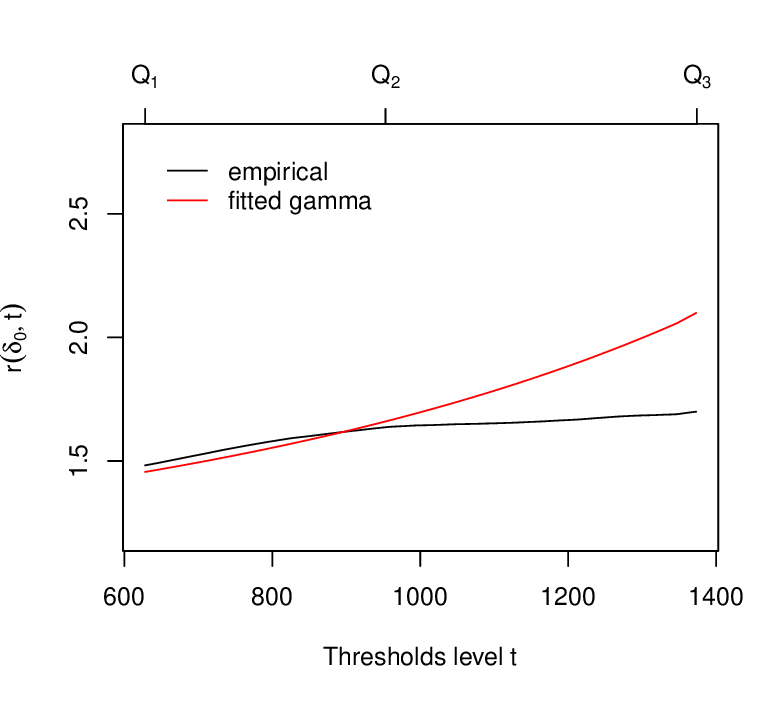}}
\vspace{2.5ex}
\caption{Values of the ratios   $r(\delta_0,t)$ and $\hat r(\delta_0,t)$ for the fire loss data,  where  $Q_1$, $Q_2$ and $Q_3$ represent the $1$st, $2$nd and $3$rd quartiles of the data} \label{fig6}
\end{figure}

We are interested in   the ratio $r(t,\delta)$ of 
the worst-case mean excess loss to that of the benchmark distribution, 
defined by 
$$ r(\delta, t) = \frac{ \sup \{  \E[(Y-t)_+]: W_{2}(F_X, F_Y) \leq \delta \}   }{ \E[(X-t)_+]},$$
where $X$ follows one of the benchmark distributions (3 choices for each dataset). We first fix the threshold level $t $   as the first quartile (25\% quantile) $t_0$ of its corresponding benchmark distribution and let $\delta$ vary (Figures \ref{fig3} and \ref{fig4}), and then we fix $\delta=\delta_0$ and let $t$ vary (Figures \ref{fig5} and \ref{fig6}).

Figure \ref{fig3} (a) and (b) present goodness-of-fit plots for the fitted lognormal, Weibull and Gamma distributions to the hurricane data. We observe that the lognormal and Weibull distributions fit better to this dataset than the Gamma distribution. 
In Figure \ref{fig3} (c), the Gamma model is penalized more for model uncertainty. 
We note that  the curves  $\delta \mapsto r(\delta,t_0)$ are almost linear in $\delta$. The numerical values of  $r(\delta,t_0)$ in Table \ref{tab:hurr} show  that $r  (\delta,t_0)$ is actually convex in $\delta$, implying that the worst-case mean excess loss becomes more sensitive to $\delta$ for large values of $\delta$,  consistent with the numerical analysis in Section \ref{sec:app}.  
Figure \ref{fig4} based on  the fire loss data exhibits    similar patterns to the hurricane loss data. The lognormal distribution fits better to the fire loss than other two distributions so that the mean excess loss will be less affected by the Wasserstein uncertainty. 

\begin{table}[t]
\begin{center}
\begin{tabular}{c|c|cccccc}
\hline
\hline
 & & $\delta_0$ & $1.2 \delta_0$ & $1.4 \delta_0$ & $1.6 \delta_0$ & $1.8 \delta_0$ & $2 \delta_0$ \\
\hline
\multirow{3}{*}{Hurricane loss}   &  Lognormal& 1.708 & 1.839 & 1.985 & 2.132 & 2.279 & 2.425 \\
 &  Weibull& 1.853 & 2.012 & 2.193 & 2.352 & 2.534 & 2.715 \\
 &  Gamma&  1.964 & 2.149 & 2.334 & 2.539 & 2.724 & 2.950 \\
\hline
\hline
\multirow{3}{*}{Fire loss} &  Lognormal& 1.358 & 1.431 & 1.505 & 1.582 & 1.657 & 1.735 \\
 &  Weibull& 1.400 & 1.481 & 1.564 & 1.649 & 1.733 & 1.819 \\
 &  Gamma&  1.456 & 1.548 & 1.644 & 1.740 & 1.837 & 1.937 \\
\hline
\hline
\end{tabular}
\end{center}
\caption{Values of  $r(\delta,t_0)$ for  the hurricane loss and the fire loss datasets. The  threshold level $t_0$ is  the first quartile of the benchmark distribution and the  parameter $\delta_0$ is the Wasserstein metric between the empirical   and the fitted distributions}\label{tab:hurr}
\end{table}


Comparing the curves $r(\delta,t_0)$ for two datasets, we can notice that the values of  $r(\delta,t_0)$ for the hurricane data are much higher than the ones for the fire data, which means the hurricane loss is more severely affected by model uncertainty than fire loss. It may be explained by the fact that the hurricane losses are more catastrophic and right-skewed than fire losses so that more penalties should be added to hurricane case if model uncertainty is a concern.

In an insurance pricing context, the mean excess loss can be used to price the stop-loss premium where the threshold $t$ can be seen as a  deductible, and our method can be used to analyze the sensitivity of the stop-loss premium to the Wasserstein uncertainty.
Taking the lognormal distribution as an example in Table \ref{tab:hurr}, if we use $\delta=\delta_0$ and $t=t_0$ for pricing a hurricane insurance, the stop-loss premium will increase 70.8\% compared to the one without considering model uncertainty. For the same choice  $\delta=\delta_0$ and $t=t_0$  (although both $t_0$ and $\delta_0$  depend on the dataset), the stop-loss premium will only increase 35.8\% when pricing the fire insurance. It intuitively means that the hurricane insurance pricing is more sensitive to the Wasserstein uncertainty than the fire insurance pricing.  

Next, we investigate how different threshold levels $t$ may influence the mean excess loss with and without the Wasserstein   uncertainty. The uncertainty level $\delta$ is fixed as $\delta_0$ in this experiment and we look at $r(\delta_0,t)$ as $t$ varies. 
Figures \ref{fig5} and \ref{fig6} report  the ratio $r(\delta_0,t)$  in these settings, as well as the ratio 
$$ \hat r(\delta_0, t) = \frac{ \sup \{  \E[(Y-t)_+]: W_{2}(F_X, F_Y) \leq \delta_0 \}   }{ \E[(\hat X-t)_+]},$$
 where $\hat X$ follows the empirical distribution of the data. 
Note that  $\hat r(\delta_0,t)\ge 1$ since $\delta_0$ is chosen such that the empirical distribution is inside the Wasserstein ball. 
For both datasets, we let the threshold level $t$ vary between the first quartile and the third quartile of the loss data. We observe from Figure \ref{fig5} that the ratio $r(\delta_0,t)$  for the hurricane loss data is relatively stable with respect to the threshold level $t$, whereas Figure \ref{fig6} shows that the ratio $r(\delta_0,t)$ for the fire loss data increases with the threshold level $t$ in all selected benchmark distributions.  Hence, compared to hurricane loss, the mean excess loss of fire loss data is more sensitive to model uncertainty with larger threshold levels. This observation is less pronounced for the better fitted lognormal distribution in Figure \ref{fig6} (a). The other ratio  $\hat r(\delta_0, t)$  is quite stable for the fire loss data and it shows a decreasing trend in $t$ for the hurricane loss data.

\section{Optimized certainty equivalents}\label{sec:OCE}

We proceed to offer some more theoretical results and discussions on the reverse ES optimization formula. 
It would be interesting to see whether Theorem \ref{th:new1} can be generalized to other risk measures than the class $\ES_\alpha$. 
Note that $\ES_\alpha$ belongs to the class of optimized certainty equivalents (OCE) of \cite{BT07}.
The class of OCE includes  ES and the entropic risk measures (\cite{FS16}) as special cases.
In this section, we work with    the set $\X_B$ of essentially bounded random variables to avoid integrability issues. 
Let $V$ be the set of   increasing and convex functions $v:\R\to \R$  satisfying $v(0)=0$,   $\bar v \ge 1$ and $\lim_{t\to\infty} v'_+(-t)=0$ where  $\bar v =\sup_{x\in \R} v'_+(x) $ and $v'_+$  is the right derivative of $v$. 
An OCE is a risk measure $R$  defined by 
$$
R(X) = \inf_{t\in \R}\left\{t  + \E[v(X-t)]\right\},~~~~X\in \X_B.
$$
 The finiteness of $R$ is guaranteed if $v'_+(x)\le 1\le v'_+(y)$ for some $x,y\in \R$ which is satisfied by $v\in V$ if $\bar v>1$. 
 If $R$ is finite, then it is a convex risk measure in the sense of \cite{FS16}. 
 In particular, if $v(x) = x_+/(1-\alpha)$ for $\alpha \in [0,1)$, then $R$ is $\ES_\alpha$ as in Theorem \ref{th:ru02}. Moreover, under a continuity condition, $\ES_\alpha$ is the only class of coherent risk measures in the class of OCE (Theorem 3.1 of \cite{EMWW21}). 


Inspired by Theorem \ref{th:new1}, we define a parametric family of OCE risk measures. For $v\in V$   and $\beta\in (0, \bar v]$, let
$$
R^v_{\beta} (X)= \inf_{t\in \R}\left\{t  + \frac{1}{\beta}\E[v(X-t)]\right\},~~~~X\in \X_B.
$$  
Here, the convention is $1/\infty=0$.
If $v(x)=x_+$, then $\bar v=1$ and $R^v_{\beta}=\ES_{1-\beta}$ for $\beta \in (0,1]$. 
The next result gives a reverse optimization formula for OCE, which  
 includes the formula \eqref{eq:new1} as a special case.
This result is related to the Fenchel-Legendre transformation as we discuss in Section \ref{sec:FL}.

\begin{theorem}[Reverse OCE optimization formula]
\label{th:new2} For $X\in \X_B$, $t\in \R$ and $v\in V$, it holds
\begin{align} \label{eq:OCE1} \E[v(X-t)  ]  =  \sup_{\beta \in (0,\bar v ]}\left\{  \beta  ( R^v_\beta(X) -t  ) \right\}.
 \end{align} 
 \end{theorem}
  \begin{proof}
  Let $f: \R\to \R$ be defined by $f(t)= \E[v(X+t)]$, which is an increasing convex function on $\R$. As a convex function on $\R$, $f$ is automatically continuous. 
Its  Fenchel-Legendre transform $f^*$, called the conjugate function of $f$,  is given by
\begin{equation}\label{eq:conjugate}
f^{*}(\beta) = \sup_{t\in \R} \left\{t \beta  - f(t)\right\}, \quad \beta \in \R,
\end{equation}
which is not necessarily finite. 

If $\beta<0$, then letting $t\to-\infty$ gives $\sup _{t\in \R}\{ t \beta -f(t)\}=\infty$ since $f$ is increasing.   
On the other hand, if $\beta>\bar v$, then, since $v_+'(x)\le \bar v$ for $x \in \R$,  letting $t\to\infty$ gives $\sup _{t\in \R}\{ t \beta -f(t)\}=\infty$. 

 Let $s\in \R$ be such that $f'_+(s)>\bar v/2$; such $s$ exists since $\lim_{t\to\infty}f_+'(t)=\bar v$. For $\beta\in (0,\bar v/2]$, we have 
$$
f^{*}(\beta) = \sup_{t\in (-\infty,s]} \left\{t \beta  - f(t)\right\} \le s \beta +\sup_{t\in \R} \{-f(t)\} = s\beta + f^*(0).
$$ 
Hence, \begin{equation}
\label{eq:FL2} \limsup_{\beta\downarrow 0} f^*(\beta)\le f^*(0) .
\end{equation}
Summarizing the above observations, for a fixed $t\in \R$,
\begin{equation}
\label{eq:FL1}
 \sup_{\beta \in \R} \left\{-t \beta  - f^*(\beta)\right\} = \sup_{\beta \in [0,\bar v]} \left\{-t \beta  - f^*(\beta)\right\}= \sup_{\beta \in (0,\bar v]} \left\{-t \beta  - f^*(\beta)\right\},
 \end{equation}
 where the last equality is due to \eqref{eq:FL2}.
For $\beta \in (0,\bar v]$,
$$  \frac{ - f^*(\beta)}{\beta} = \inf_{t\in \R} \left\{- t   + \frac{1}{\beta}f( t)\right\} =\inf_{t\in \R} \left\{ t   + \frac{1}{\beta}f(-t)\right\} = R^v_\beta(X).
$$ 
Thus, $f^*(\beta) = - \beta  R^v_{\beta}(X)$. 
The Fenchel-Legendre theorem in the form of Proposition A.9 of \cite{FS16} gives $f^{**}=f$. Therefore, using \eqref{eq:FL1}, 
\begin{align*}
f(-t) & = \sup_{\beta \in (0,\bar v]} \left\{-t \beta  - f^*(\beta)\right\}
 = \sup_{\beta \in (0,\bar v] } \left\{-t \beta  + \beta  R^v_{\beta}(X)\right\}
 = \sup_{\beta \in (0,\bar v]} \left\{  \beta   (R^v_{\beta}(X)-t)\right\},
\end{align*} 
and hence \eqref{eq:OCE1} holds.
 \end{proof}
 
 As we can see from Theorem \ref{th:new2}, the symmetry between the ES optimization formula \eqref{eq:old} and the reverse formula \eqref{eq:new1} can be seen as a consequence of the Fenchel-Legendre transform mechanism.
 
\section{Related Fenchel-Legendre transforms}\label{sec:FL}
As mentioned above, the reverse ES optimization formula and reverse OCE optimization formula is closely related to the Fenchel-Legendre transformation (e.g., Definition A.8 of \cite{FS16}). In this section, we give two pairs of conjugate functions related to Theorem \ref{th:new1}.

The Fenchel-Legendre transformation converts convex functions to their conjugate. For a convex function $f: \R \to \R$, its Legendre-Fenchel transform is the function $f^{*}$ on $\R$ defined by
$$ f^{*}(\beta) = \sup_{t\in\R} \left\{t \beta - f(t)\right\},~~~~\beta \in \R,$$
where $\beta$ may be constrained to a subset of $\R$ such that $f^*$ is real. 

%

As we have seen in Theorem \ref{th:new2},  Fenchel-Legendre transforms are  closely related to our reverse ES optimization formula, as Theorems \ref{th:ru02} and \ref{th:new1} can be expressed from each other via a Fenchel-Legendre transform. 
Below, we identify two other pairs of conjugate functions, one being quantile-based and one being expectation-based, analogously to the case of ES and the mean excess function.

\begin{proposition}
\label{th:new3} 
Fix $X \in \X$.
\begin{itemize}
\item[(i)] The Fenchel-Legendre transform of the convex quantile-based function $f_{1}: [0,1] \to \R,$ 
$$f_{1} (\alpha) = -(1-\alpha)\ES_{\alpha}(X),$$ is given by $$ f_{1}^{*}(t) = \max_{\alpha \in [0,1]} \left\{\alpha t - f_{1}(\alpha)\right\} = \E[X \vee t].$$ 
\item[(ii)]The Fenchel-Legendre transform of the convex quantile-based function $f_{2}: [0,1] \to \R,$  
$$f_{2} (\alpha) = \alpha\ES^{-}_{\alpha}(X),$$ is given by $$ f_{2}^{*}(t) = \max_{\alpha \in [0,1]} \left\{\alpha t - f_{2}(\alpha)\right\} = \E[(t-X)_{+}].$$
\end{itemize}
Moreover, the set of  maximizers for both maximization problems is $[\mathbb{P}(X < t), \mathbb{P}(X \leq t)]$.
\end{proposition}

\begin{proof}
For the first statement,  it is straightforward to identify that the quantile-based function ${f_{1} : \alpha \mapsto -(1-\alpha)\ES_{\alpha}(X)}$ is convex by taking a derivative with respect to $\alpha$. By definition of the Legendre-Fenchel transformation, we have
\begin{align*}
f_{1}^{*}(t) &= \sup_{\alpha \in [0,1]} \left\{\alpha t + (1-\alpha)\ES_{\alpha}(X)\right\}\\
&= \sup_{\alpha \in [0,1]} \left\{(\alpha-1) t + (1-\alpha)\ES_{\alpha}(X)\right\} +t\\
&= \sup_{\alpha \in [0,1]} \left\{(1-\alpha)(\ES_{\alpha}(X)-t)\right\} +t.
\end{align*}
By the reverse ES optimization formula  in Theorem \ref{th:new1}, we know that ${\alpha \in [\P(X<t), \P(X\leq t)]}$ is a maximizer of function $\alpha \mapsto (1-\alpha (\ES_{\alpha}(X) - t))$, and hence the supremum above is attainable. Thus, we can conclude that
$$f_{1}^{*}(t) = \E[(X-t)_{+}] + t = \E[X \vee t].$$ 
The proof of the second Fenchel-Legendre transform follows the same routine. We apply the Fenchel-Legendre transform to the convex function $f_{2}(\alpha) = \alpha \ES^{-}_{\alpha}(X) .$ Then we have
\begin{equation*}
\begin{aligned}
f_{2}^{*}(t) &= \sup_{\alpha \in [0,1]} \left\{\alpha t - \alpha \ES^{-}_{\alpha}(X) \right\}\\
&= \sup_{\alpha \in [0,1]} \left\{\alpha t - \E(X) + \int^{1}_{0}\VaR_{\beta}^{-}(X)\d\beta - \int^{\alpha}_{0}\VaR_{\beta}^{-}(X)\d\beta \right\}\\
&= \sup_{\alpha \in [0,1]} \left\{(1-\alpha)(\ES_{\alpha}(X)-t)\right\} +t - \E[X].\\
\end{aligned}
\end{equation*}
By Theorem \ref{th:new1}, we conclude that
$$f_{2}^{*}(t) = \E[(X-t)_{+}] + t - \E[X]= \E[(t - X)_{+}].$$
We can check that both 
 $f_{1}^{*} : t\mapsto \E[X \vee t] $ and $f_{2}^{*}:t \mapsto \E[(t - X)_{+}] $ are  convex functions.
\end{proof}
 
By applying  Legendre-Fenchel transform mechanism  to the convex functions ${f_{1}^{*}:t\mapsto   \E[X \vee t]}$ and $f^{*}_{2}:t\mapsto  \E[(t - X)_{+}]$, one obtains the corresponding quantile-based  functions $f_1$ and $f_2$  in
Proposition \ref{th:new3} as their conjugate functions.

\section{Conclusion}
\label{sec:conclusion}
The reverse ES optimization formula obtained in Theorem \ref{th:new1} serves as a dual formula to the celebrated ES optimization formula of \cite{RU00, RU02}, and they are connected via the Fenchel-Legendre transforms. This new formula reveals profound symmetries between these two formulas regarding to their functional properties, parametric forms, optimization problems and the solutions to the optimization problems,
and it can be generalized for the class of  OCE of \cite{BT07}.
The reverse ES optimization formula is particularly useful when directly calculating the mean excess loss is cumbersome, and this is illustrated by two settings of model uncertainty.
The new formulas are applied to settings of model uncertainty and two insurance datasets.

The new formula may appear   simple to risk experts, although we could not find it in the literature.
The reason why such a natural formula has not been studied could partially be explained by the fact that
the need for utilizing existing ES results to compute the mean excess loss 
mainly arises in the recent years, when model uncertainty is actively studied in quantitative risk management, as we discuss in Sections \ref{sec:app} and \ref{sec:5}.

The main purpose of this  paper is to introduce the new formula and discuss its direct implications.    
Given the importance of both the mean excess function and ES in actuarial science and risk management, we are optimistic about other potential applications of the formula, which will need future exploration.

\subsection*{Acknowledgements}
Ruodu Wang acknowledges financial support from the Natural Sciences and Engineering Research Council of Canada (RGPIN-2018-03823, RGPAS-2018-522590).

\end{document}